\documentclass[floats,floatfix,showpacs,amssymb,prl,twocolumn,superscriptaddress,nofootinbib]{revtex4-2}

\usepackage{graphicx}
\usepackage{dcolumn}
\usepackage{bm}
\usepackage{booktabs}
\usepackage{hyperref}
\usepackage{makecell}
\usepackage{amsmath}
\usepackage{xcolor}

\hypersetup{
  colorlinks=true,
  linkcolor=black,
  citecolor=black,
  urlcolor=black
}

\begin{document}

\title{Identifying the host of compact binary mergers}

\author{Alberto Salvarese}
\affiliation{Department of Physics, The University of Texas at Austin, 2515 Speedway, Austin, TX 78712, USA}
\email{alberto.salvarese@utexas.edu}
\author{Hsin-Yu Chen}
\affiliation{Department of Physics, The University of Texas at Austin, 2515 Speedway, Austin, TX 78712, USA}
\author{Daniel E. Holz}
\affiliation{Department of Physics, University of Chicago, Chicago, IL 60637, USA}

\begin{abstract}
Finding the host galaxies of stellar-mass compact binary mergers will open a new window for studying their formation histories and measuring key cosmological parameters, such as the Hubble constant. To date, only one merger, GW170817, has had its host galaxy confidently identified through electromagnetic counterpart observations. The large localization volumes from the LIGO--Virgo--KAGRA (LVK) network, combined with the lack of electromagnetic emission for most events, make host identification challenging. However, as the sensitivity of the gravitational-wave (GW) detector network improves, events are becoming increasingly well localized. Furthermore, galaxy luminosity traces mass or star formation rate, and thus correlates with the probability of hosting a merger. Focusing on the most luminous galaxies within the localization volumes of the best-localized GW events, we estimate the corresponding Hubble constant for each galaxy by combining its redshift with the luminosity distance inferred from LVK observations.
For the well-localized LVK events \texttt{S250207bg}, \texttt{GW190814}, and \texttt{S250830bp}, we find only 1, 1, and 4 galaxies, respectively, when restricting the analysis to the most luminous 1\% of galaxies above $L_{\rm th}\sim 10^{9} h^{-2}L_{\odot}$ in each event's localization volume and adopting a broad $H_0$ prior.
The probability of these galaxies being random, and not associated with the GW events, is 29--36\% across the three events. We encourage further follow-up observations of these candidate host galaxies. We expect this approach to become increasingly powerful in future LVK observing runs, enabling constraints on merger formation histories and measurements of the Hubble constant.
\end{abstract}

\maketitle

\section{Introduction}
Population-synthesis and cosmological models predict that stellar-mass compact binary mergers (CBCs) may preferentially occur in galaxies with specific stellar masses, metallicities, star-formation histories, or morphologies (e.g., \cite{Cao_2017, Artale_2019doq, 2020ApJ...905...21A, Mapelli_2020, Zevin_2021, Santoliquido_2022, Vijaykumar:2023bgs}). Identifying the host galaxies of CBCs therefore provides a unique opportunity to investigate the astrophysical environments in which compact binaries form and merge, and to probe their formation histories. Moreover, by combining the host-galaxy redshift with the luminosity distance measured from LVK GW observations, we can constrain cosmological parameters using the standard-siren method (e.g., \cite{Schutz1986, Holz_2005, Taylor_2012, DelPozzo2012, Chen_2018, Gair_2023, Chen_2024, Salvarese:2025qel, 2025ApJ...979....9H}), providing an independent probe of the cosmic expansion rate.

For the current generation of GW detectors, however, the identification of a unique host galaxy for CBCs is highly unlikely, due to the relatively large sky-localization areas and distance uncertainties associated with GW measurements. Even for well-localized events, the three-dimensional GW localization volume typically encompasses thousands of galaxies. While events with an electromagnetic (EM) counterpart allow for a direct host-galaxy association, such events are extremely rare, with only a single confirmed case to date \citep[]{GW170817_Observation,GW170817_2017}.
\begin{table*}[]
    \centering
    \caption{The five best-localized events in O3 and O4 and their properties: GraceDB 90\% sky area and distance, estimated comoving volume, and number of galaxies to retain $N_g$.}
   \label{tab:gds_volume}
    \begin{tabular}{ccccc}
    \toprule
    & $90\%$ Sky area [deg$^2$] & Distance [Mpc] & Volume [$\times10^3$ Mpc$^3$] & $N_g$\\ 
    \midrule
    \href{https://gracedb.ligo.org/superevents/S250207bg/view/}{S250207bg} & 19 & $180\pm38$ & 18.1 & 2\\
    \href{https://gracedb.ligo.org/superevents/S190814bv/view/}{GW190814} & 19 & $241\pm26$ & 23.9 & 2\\
    \href{https://gracedb.ligo.org/superevents/S250830bp/view/}{S250830bp} & 3.73 & $427\pm69$ & 33.0 & 3\\
    \href{https://gracedb.ligo.org/superevents/S240925n/view/}{S240925n} & 14 & $331\pm70$ & 76.9 & 8\\
    \href{https://gracedb.ligo.org/superevents/S241011k/view/}{S241011k} & 72 & $212\pm37$ & 99.7 & 10\\
    \bottomrule
    \end{tabular}
\end{table*}

Recent studies demonstrated that targeting the most luminous galaxies in a catalog can preserve unbiased cosmological inference, despite discarding the majority of potential hosts (e.g., \cite{Bera_2020, naveed2025darkstandardsirencosmology, Vanwyngarden_2025, Ghosh_2026}). This result is physically motivated by the fact that luminous galaxies trace stellar mass or star formation rate and could be embedded in groups and clusters populated by fainter satellite galaxies. 
In particular, redder bands are expected to more closely trace stellar mass, whereas bluer bands more directly trace star formation rate. The most luminous galaxies can therefore serve as effective proxies for the host location, even when the true host galaxy is not directly identified.

The large number of candidate galaxies and the incompleteness of available galaxy catalogs for typical GW events can limit the effectiveness of this approach. However, the best-localized GW events have significantly reduced false identification rate \citep[]{chen2014loudestgravitationalwaveevents, chen2016findingoneidentifyinghost}, as we will show below.

Focusing on the most luminous galaxies within the localization volume of the best-localized GW events, we assess whether they support a common range of values of the Hubble constant, $H_0$. Galaxies that do so are identified as potential host galaxies or as members of the host cluster of the GW event.

\section{Method}
\noindent\textbf{{\em Brightest galaxies within the best-localized events.---}}
We start by identifying the best-localized GW events in the LVK third and fourth observing runs (O3 and O4; \cite{quantum_enhanced, Buikema_2020, Capote_2025, Soni_2025}).  For each event, we compute its comoving volume using its corresponding GraceDB sky map (\href{https://gracedb.ligo.org/}{https://gracedb.ligo.org/}), which provides a conditional luminosity distance posterior along each line-of-sight (LOS). We consider the 90\% credible region in the sky and the symmetric 90\% credible interval of the luminosity-distance posterior (for details, see the Appendix). We adopt a fiducial flat $\Lambda$CDM cosmology (e.g., \cite{Dodelson:2003ft, Lyth:2009imm, weinberg_2008zzc}) with $H_0^{\rm fid} = 67.74\ \mathrm{km\,s^{-1}\,Mpc^{-1}}$
and $\Omega_m^{\rm fid} = 0.3075$ \citep[]{Planck_2015}. In Table~\ref{tab:gds_volume}, we report the five best-localized GW events' sky area, luminosity distance, and comoving volume.

With the public \textsc{GLADE+} galaxy catalog \citep[]{GLADE+} \footnote{\textsc{GLADE+} galaxy catalog provides sky coordinates---right ascension and declination $(\alpha, \delta)$---redshift information, and multi-band photometry for approximately $22.5$ million galaxies.}, we examine the spatial distribution of the galaxies within the localization volume of each GW event. We find that only \texttt{S250207bg}, \texttt{GW190814}, and \texttt{S250830bp} show an approximately spatially uniform galaxy distribution. Therefore, we restrict our analysis to these three events. The non-uniform distributions for the remaining events are primarily due to their sky localizations overlapping with or lying near the Galactic plane, where incompleteness and extinction in galaxy catalogs significantly affect the observed galaxy density. An illustrative example of a discarded event is shown in the Appendix.

For each GW event, we determine the number of most luminous (i.e., lowest absolute magnitude) galaxies to include, $N_g$, as $N_g = \rho_gV_{\rm C}f_g$, where $V_{\rm C}$ is the event comoving localization volume listed in Table~\ref{tab:gds_volume}. We adopt a luminosity threshold $L_{\rm th}\sim 0.4$--$7 \times 10^{9} h^{-2} L_{\odot}$, depending on the photometric band ($B$, $B_J$, $J$, $K$, $H$, $W1$, and $W2$), such that the mean galaxy number density $\rho_g = 10^{-2}{\rm Mpc}^{-3}$ ~\footnote{{$L_{\rm th}$ is computed by integrating the Schechter function \citep{Schechter:1976iz} for the different considered bands and requiring a cumulative number density of $\rho_g\sim 10^{-2}{\rm Mpc}^{-3}$ \citep{b_band, bj_band, j_band, H_band, k_band, W1W2_band}}.} for $h=0.6774$. With this choice, we consider the same $N_g$ across different photometric bands for a fixed fraction $f_g$, which we take to be $f_g=1\%$. The resulting values of $N_g$ are reported for each event in Table~\ref{tab:gds_volume}. We discuss these choices further in the Discussion.

We neglect the K-correction since all considered events are in the local universe, where its effect is expected to be small compared to other sources of uncertainty (e.g., \cite{Chilingarian_2010, Hui_2006}). We note that adopting a different fiducial $H_0$ would shift the absolute magnitude values of individual galaxies, but would not affect their relative ranking in magnitude, and therefore would not alter the selection of the most luminous galaxies. 

\noindent\textbf{{\em $H_0$ consistency.---}}
We estimate the $H_0$ posterior for each considered galaxy by combining their redshifts and the luminosity-distance likelihood along the LOS from LVK observations (see the Appendix).

With the $H_0$ posterior, we identify galaxies that satisfy the following two conditions simultaneously. Let $q_{H_0}^{x\%}$ denote the $x\%$ quantile of the $H_0$ posterior: 
\begin{itemize}
    \item \textit{$68\%$ credible interval support}: $q_{H_0}^{84\%} \geq 60\,\mathrm{km\,s^{-1}\,Mpc^{-1}}$ \quad \text{and}\quad $q_{H_0}^{16\%} \leq 80\ \mathrm{km\,s^{-1}\,Mpc^{-1}}$;
    \item \textit{Median support}: $q_{H_0}^{50\%} \in [50, 90]\, \mathrm{km\,s^{-1}\,Mpc^{-1}}$.
\end{itemize}
The $H_0$ range was chosen to be broadly consistent with current observational constraints (e.g., \cite{Planck:2018vyg, h0dncollaboration2025localdistancenetworkcommunity})\footnote{Our results remain unchanged for a narrower choice of the $H_0$ range. On the other hand, adopting a wider $H_0$ range would increase the probability of random associations (see next section) and is not necessary given current observational constraints on $H_0$.}.

\begin{table*}[ht!]
    \begin{center}
    \caption{Identified galaxies for each GW event. The table reports the galaxies identified in our analysis together with their main properties: name and catalog, redshift $z$, right ascension $\alpha$, declination $\delta$, and the magnitude and band in which the galaxy was identified.}
    \label{tab:galaxies_with_support}
        \begin{tabular}{ccccccc}
        \toprule
        \makecell[c]{Name and catalog} & $z$ & $\alpha \, [^\circ]$ & $\delta \, [^\circ]$ & \makecell[c]{Magnitude} & $H_0 [{\rm km}\,{\rm s}^{-1}\,{\rm Mpc}^{-1}]$ & \makecell[c]{Prob. of random\\ association} \\
        \midrule
        \multicolumn{7}{c}{\textit{S250207bg}} \\
        \midrule
        \href{https://simbad.cds.unistra.fr/simbad/sim-id?Ident=\%401726560\&Name=Z\%20183-33\&submit=submit}
        {\makecell[c]{10375180+3348504\\2MASS}} & 0.0501 & 159.47 & 33.81 & $B_J=-21.31$ & $82.29^{+12.60}_{-9.66}$ & 29.4\% \\
        \midrule
        \multicolumn{7}{c}{\textit{GW190814}} \\
        \midrule
        \href{https://simbad.cds.unistra.fr/simbad/sim-id?Ident=\%401212854\&Name=2dFGRS\%20TGS452Z029\&submit=submit}
        {\makecell[c]{100480\\HyperLEDA}} & 0.0630 & 23.59 & -32.84 & $J=-25.11$ & $83.05^{+11.29}_{-9.01}$ & 32.2\% \\
        \midrule
        \multicolumn{7}{c}{\textit{S250830bp}} \\
        \midrule
        \href{https://simbad.cds.unistra.fr/simbad/sim-id?Ident=\%405056741\&Name=LEDA\%20\%20234454\&submit=submit}
        {\makecell[c]{234454\\HyperLEDA}} & 0.0976 & 329.83 & -77.88 & $B=-21.83$ & $76.27^{+15.72}_{-11.27}$ & 36.4\% \\
        \href{https://simbad.cds.unistra.fr/simbad/sim-id?Ident=\%405055651\&Name=LEDA\%20\%20235533\&submit=submit}
        {\makecell[c]{235533\\HyperLEDA}} & 0.0715 & 324.25 & -77.57 & $B=-21.72$ & $52.08^{+9.25}_{-6.91}$ & 36.4\% \\
        \href{https://simbad.u-strasbg.fr/simbad/sim-coo?CooFrame=FK5&CooEpoch=2000&CooEqui=2000&CooDefinedFrames=none&Radius.unit=arcmin&submit=submit+query&Coord=J212943.1-775543&Radius=2.5}
        {\makecell[c]{234275\\HyperLEDA}} & 0.0813 & 322.43 & -77.93 & $B=-21.65$ & $61.63^{+11.76}_{-8.62}$ & 36.4\% \\
        \href{https://simbad.cds.unistra.fr/simbad/sim-id?Ident=\%4011146543\&Name=LEDA\%20\%20235373\&submit=submit}
        {\makecell[c]{21465750-7736463\\2MASS}} & 0.0998 & 326.74 & -77.61 & $B_J=-21.74$ & $74.52^{+18.50}_{-14.92}$ & 36.4\% \\
        \bottomrule
        \end{tabular}
    \end{center}
\end{table*}

\noindent\textbf{{\em Random association.---}}
To quantify the probability that at least one identified galaxy is a random association, we generate a mock galaxy catalog uniformly distributed in comoving volume assuming the fiducial $\Lambda {\rm CDM}$ cosmology defined above. We then randomly select $N_g$ galaxies within the localization volume of the GW events 500 times and record how often at least one of the $N_g$ galaxies satisfies the condition described above.

We note that the galaxy number density $\rho_g$ assumed for this mock catalog does not affect the results, as long as the total number of galaxies within the localization volume is significantly larger than $N_g$ so that the location of the galaxies is randomly distributed. 
We also note that it is unnecessary to generate a more realistic galaxy catalog with large-scale structure for this test. This is because the location of the large-scale structure would be random over the 500 realizations, and therefore, lead to the same fraction of random associations as those from a uniform catalog. 

In this mock catalog, the redshift uncertainties are randomly assigned by matching the fraction of spectroscopic and photometric redshift measurements available in \textsc{GLADE+} within the localization volume of each event. For spectroscopic measurements, we adopt a redshift uncertainty of $\sigma_z = 0.001$, while for photometric estimates we use $\sigma_z = 0.013(1+z)^3 \leq 0.015$ \citep[]{Gair_2023}. 
\section{Results} 

\begin{figure*}[ht!]
    \centering
    \includegraphics[width=0.8\linewidth]{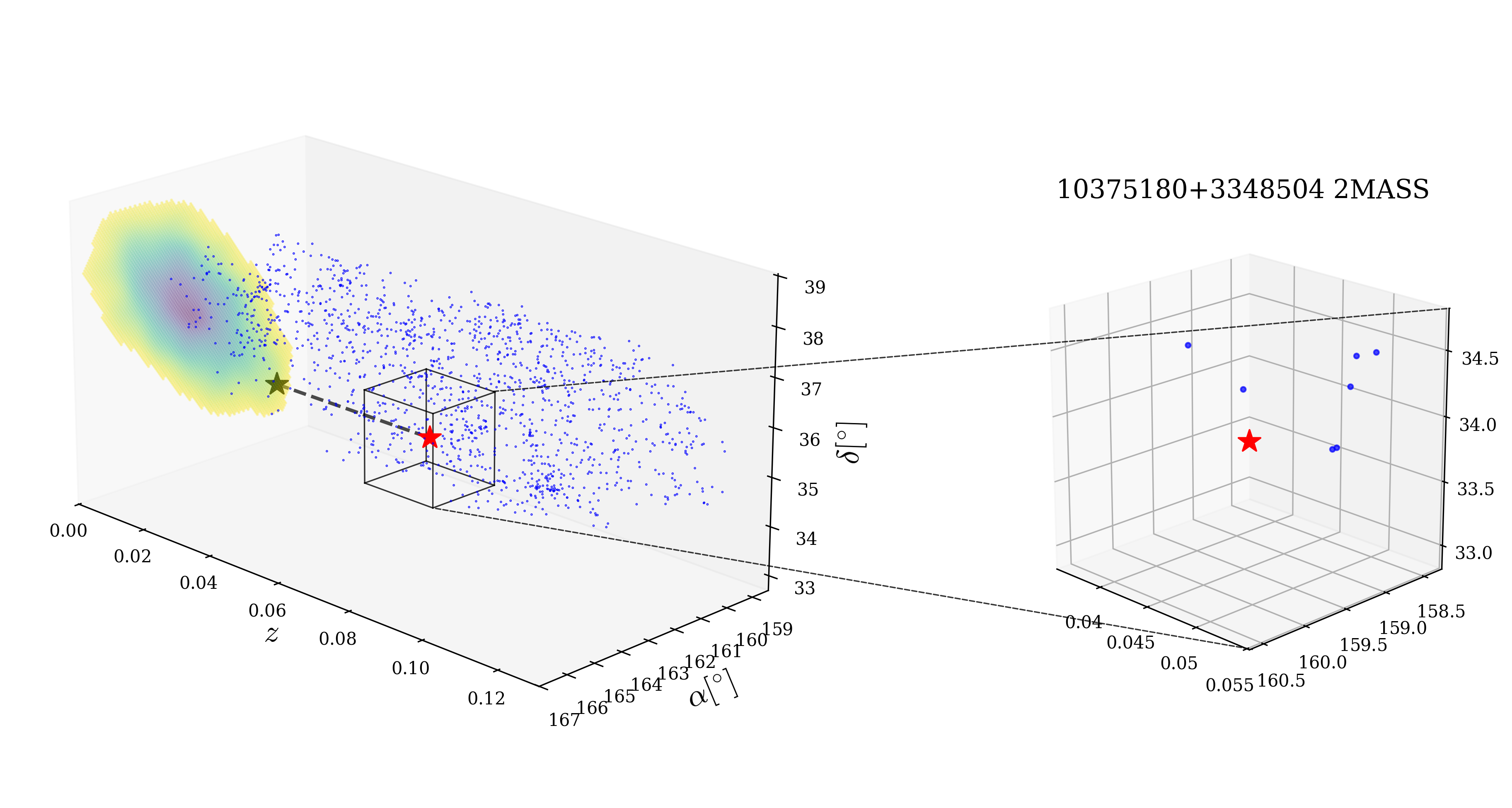}\par\vspace{-5mm}
    \includegraphics[width=0.8\linewidth]{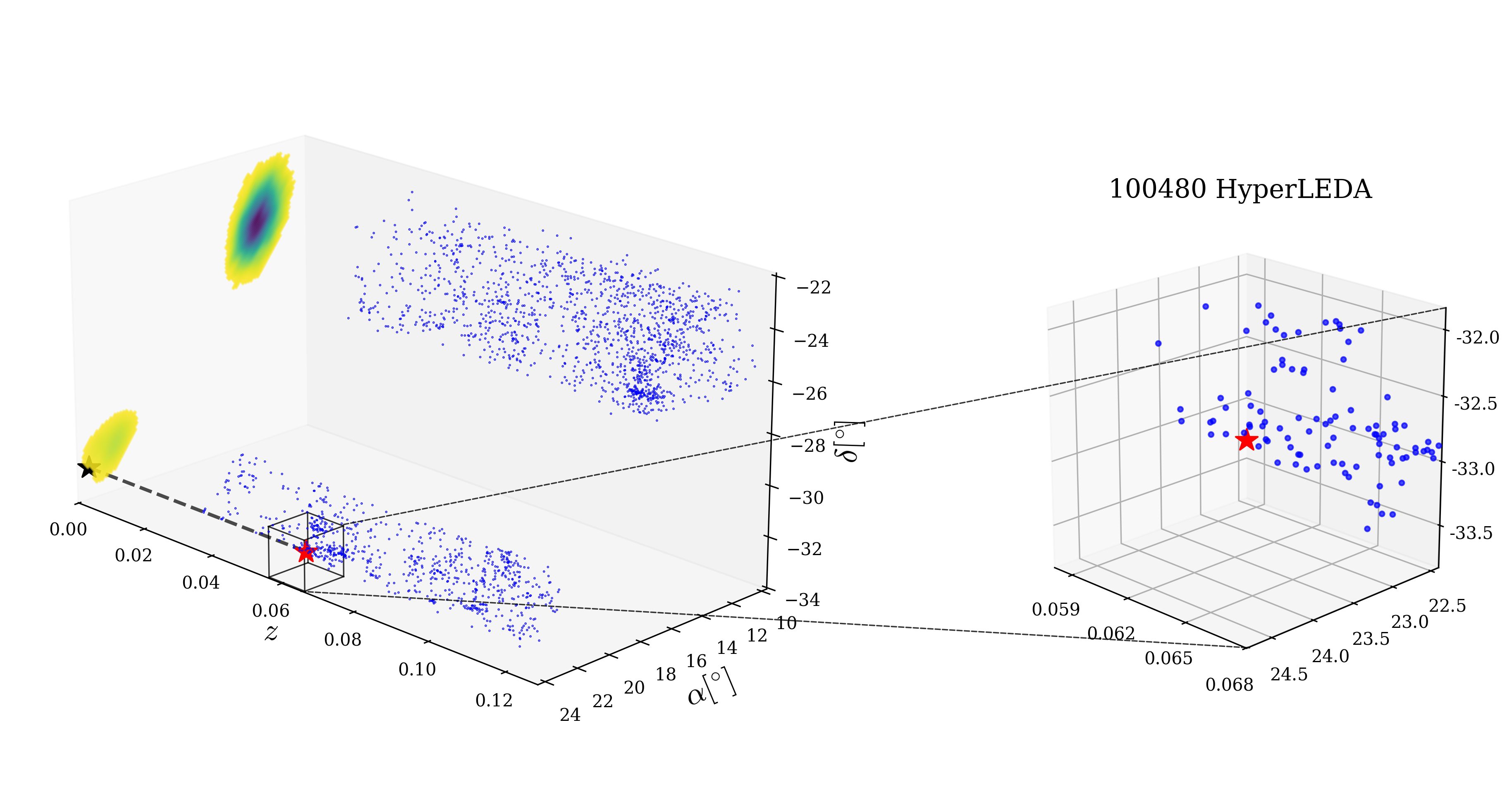}\par\vspace{-5mm}
    \includegraphics[width=0.8\linewidth]{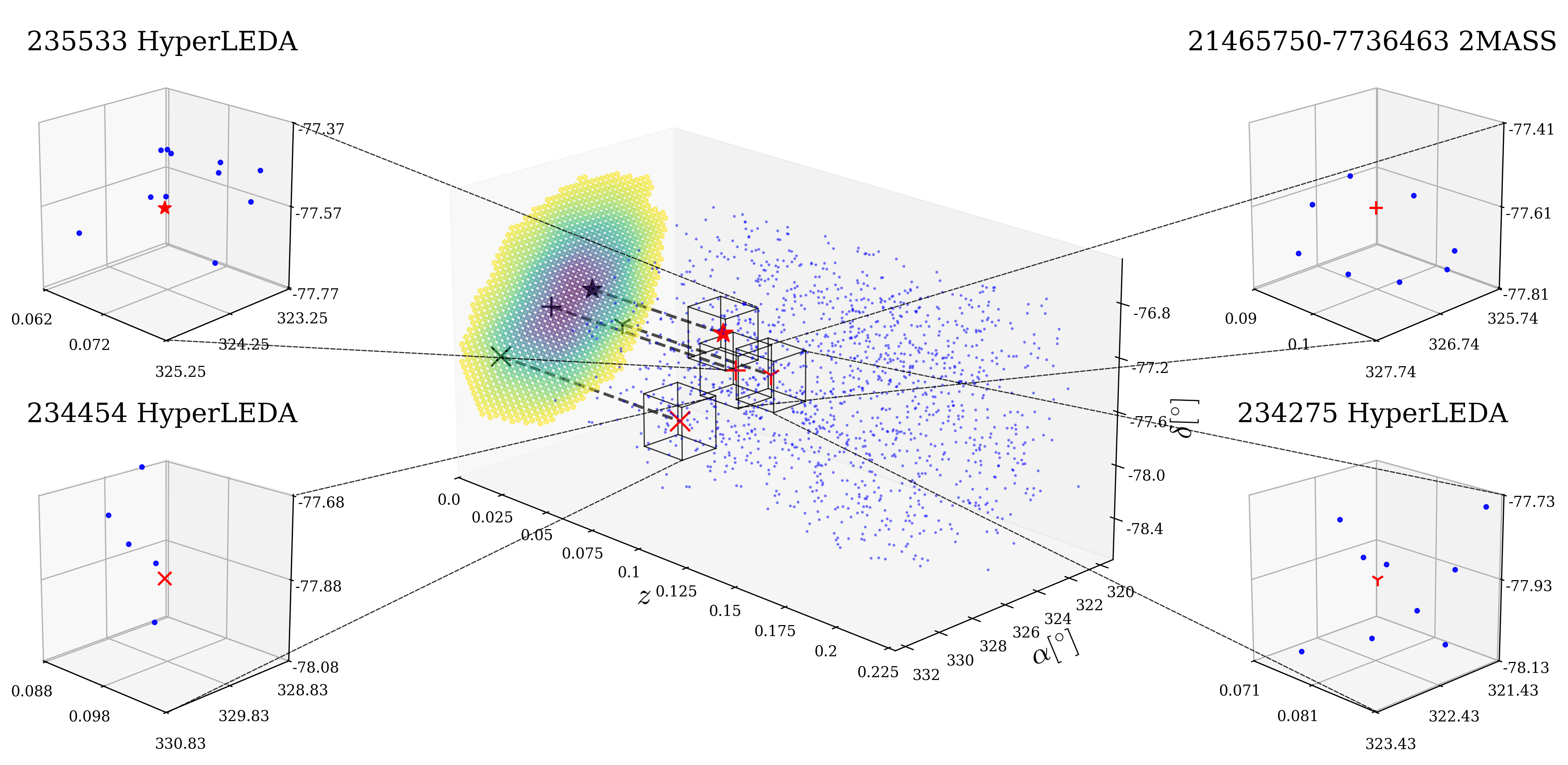}
    \caption{Three-dimensional view of the \textsc{GLADE+} galaxies (blue dots) contained within the localization volumes of \texttt{S250207bg} (top panel), \texttt{GW190814} (middle panel), and \texttt{S250830bp} (bottom panel). The sky map in each panel shows the GW 90\% credible region on the sky. The redshift extent of each volume is determined following the procedure described in the Appendix. Red symbols mark the identified galaxies reported in Table~\ref{tab:galaxies_with_support}. The zoomed-in insets highlight the local environment around each identified galaxy.}
    \label{fig:3d_events}
\end{figure*}
\noindent\textbf{{\em The identified galaxies.---}}
In Table~\ref{tab:galaxies_with_support}, we report the identified galaxies, their properties, and the median and 68\% credible interval of the $H_0$ posterior. We find 1, 1, and 4 galaxies satisfying the $H_0$-consistency conditions for \texttt{S250207bg}, \texttt{GW190814}, and \texttt{S250830bp}, respectively.
In Figure~\ref{fig:3d_events}, we show the location of the identified galaxies on the 2D sky maps of the GW events, and the 3D distribution of the galaxies in \textsc{GLADE+} around them. 

Following the morphological classifications in the \texttt{HyperLEDA} database \citep[]{HyperLEDA}, the galaxy associated with \texttt{S250207bg} is a spiral, while the galaxy associated with \texttt{GW190814} is an E-S0, a transitional type between elliptical and lenticular. The galaxy associated with \texttt{GW190814} is also marked as the brightest in a cluster. No morphological classification is available in the database for the galaxies associated with \texttt{S250830bp}.

We follow the method described above to quantify the probability of at least one random association for the identified galaxies, finding 29.4\%, 32.2\% and 36.4\% for the localization volumes of \texttt{S250207bg}, \texttt{GW190814}, and \texttt{S250830bp}, respectively. As we expected, the probability of random association increases with localization volume.

\noindent\textbf{{\em Impact of Active Galactic Nuclei.---}}
Active galactic nuclei (AGNs) can appear highly luminous due to accretion-powered emission. They may thus be preferentially included in the brightest $1\%$ sample even if their stellar content would not otherwise qualify, thereby compromising proper identification of associated galaxies in our method.

To test this, we repeat the analysis by removing all observationally confirmed AGNs from the brightest 1\% of galaxies. We apply the empirical AGN selection criteria proposed by \citet{Stern_2012} and \citet{Assef_2013}, and also check each galaxy in the \textsc{SIMBAD} database \citep[]{Wenger_2000}. We find four, one, and zero confirmed AGNs for \texttt{S250207bg}, \texttt{GW190814}, and \texttt{S250830bp}, respectively, among the brightest 1\% of galaxies considered across all photometric bands.

For \texttt{S250207bg}, the originally identified galaxy is classified as an AGN. After removing AGNs for this event, we do not find any alternative host that satisfies the $H_0$ consistency test, suggesting that the original AGN may indeed be associated with the event. For \texttt{GW190814} and \texttt{S250830bp}, the original (non-AGN) galaxies remain the only identified candidates when AGNs are excluded, also supporting their plausibility as host galaxies.
\begin{figure*}[ht!]
    \centering
    \includegraphics[width=0.95\textwidth]{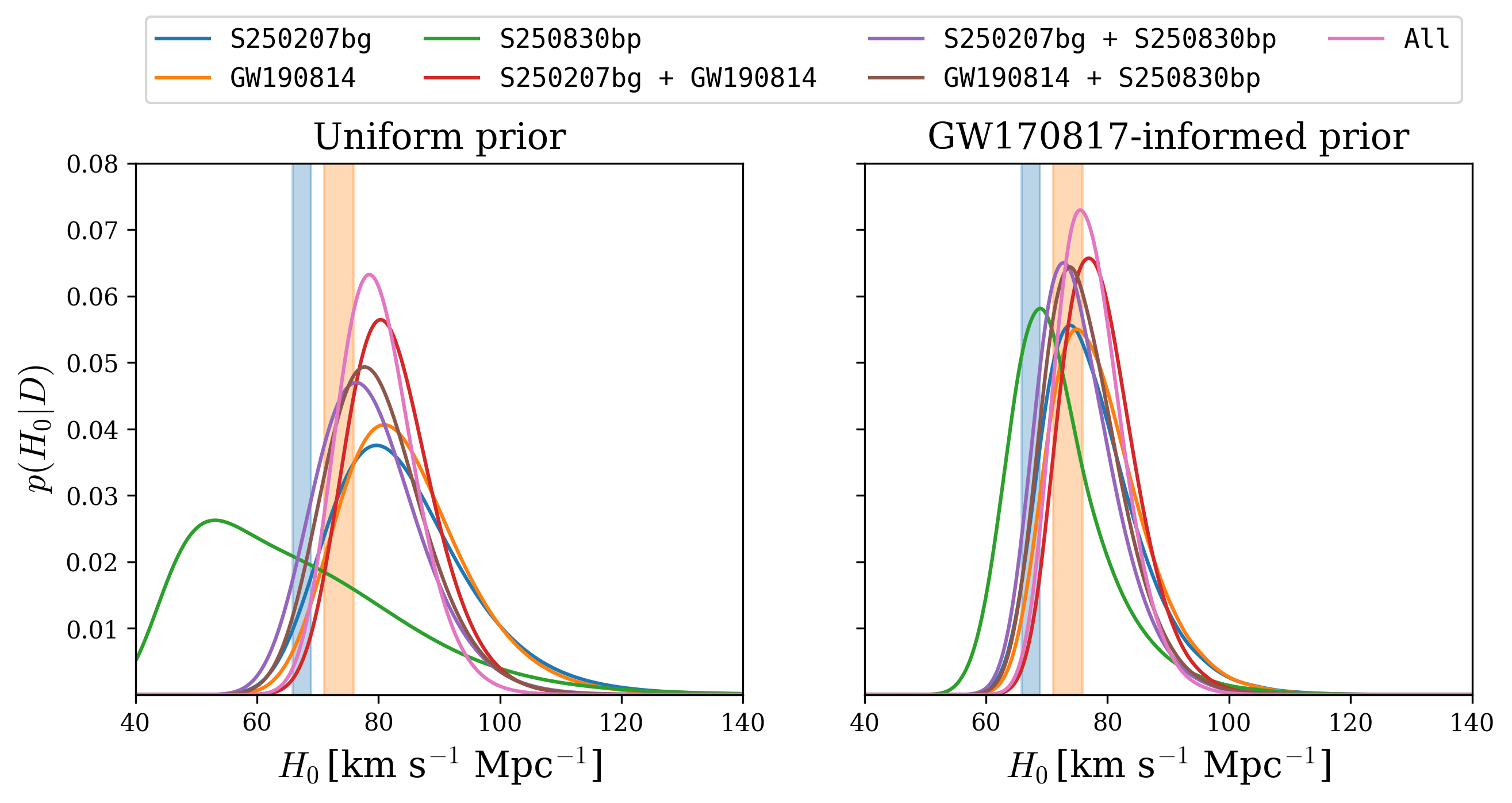}
    \caption{$H_0$ posteriors for \texttt{S250207bg}, \texttt{GW190814}, \texttt{S250830bp}, and their combinations. The left panel assumes a uniform prior $H_0\in [40,140]$ km s$^{-1}$Mpc$^{-1}$, while the right panel utilizes the $H_0$ posterior from \texttt{GW170817} as prior. The vertical shaded regions indicate the $3\sigma$ constraints on $H_0$ reported by \citet{Planck:2018vyg} (blue), and \citet{h0dncollaboration2025localdistancenetworkcommunity} (orange).}
\label{fig:hubble_constant_final_posteriors}
\end{figure*}

\noindent\textbf{{\em Hubble constant.---}}
If the identified galaxies are the true host galaxies or belong to the same clusters as the true hosts, we can use their redshift information to constrain $H_0$. When more than one galaxy is identified (e.g., \texttt{S250830bp}), we weight each galaxy by the probability of its associated LOS from GW localization. 
In Figure~\ref{fig:hubble_constant_final_posteriors} and Table~\ref{tab:H0_values}, we show the $H_0$ posterior for each GW event individually and for their combinations. In each case, we adopt two different priors: a uniform prior and a prior given by the $H_0$ posterior of \texttt{GW170817} \citep[]{GW170817_2017}. 

\section{Discussion}
In this paper we aim to identify the most plausible host galaxy of a well-localized GW event. Our approach exploits the fact that the most luminous galaxies act as effective proxies for the location of the true host. We use the Hubble constant as a consistency variable, such that agreement in the redshift-distance relation across a small set of luminous galaxies provides a physically motivated criterion to isolate the host within a GW localization volume.
\begin{table}
\begin{center}
\caption{Median and symmetric 68\% credible intervals of the  $H_0$ posterior (in ${\rm km}\,{\rm s}^{-1}\,{\rm Mpc}^{-1}$) for the individual GW events and their combinations.  Results are reported assuming a uniform prior or a \texttt{GW170817}-informed prior.}
    \label{tab:H0_values}
    \begin{tabular}{c c c}
    \toprule
    Event & Uniform prior & \texttt{GW170817}-informed prior\\
    \midrule
    \\
    \texttt{S250207bg} & $82.29^{+12.56}_{-9.65}$ & $76.31^{+9.00}_{-6.44}$\\
    \\
    \texttt{GW190814} & $83.04^{+11.28}_{-9.01}$ & $77.43^{+8.69}_{-6.51}$\\
    \\
    \texttt{S250830bp} & $63.62^{+20.28}_{-13.75}$ & $70.64^{+9.07}_{-6.26}$\\
    \\
    \texttt{\makecell[c]{S250207bg + \\ GW190814}} & $81.45^{+7.77}_{-6.58}$ & $78.09^{+6.74}_{-5.52}$\\
    \\
    \texttt{\makecell[c]{S250207bg + \\ S250830bp}} & $77.71^{+9.41}_{-7.86}$ & $74.33^{+7.34}_{-5.56}$\\
    \\
    \texttt{\makecell[c]{GW190814 + \\ S250830bp}} & $78.94^{+8.87}_{-7.54}$ & $75.38^{+7.25}_{-5.63}$\\
    \\
    \texttt{All} & $79.27^{+6.80}_{-5.92}$ & $76.65^{+6.03}_{-5.01}$\\
    \\
     \texttt{GW170817}\citep[]{GW170817_2017} & $70.0^{+12.0}_{-8.0}$ & N/A\\
     \bottomrule
    \end{tabular}
\end{center}
\end{table}

We apply our method to the three best-localized GW events detected thus far: \texttt{S250207bg}, \texttt{GW190814}, and \texttt{S250830bp}. Focusing on the most luminous $1\%$ of galaxies in each photometric band, we find that only a small number of galaxies support a reasonably broad range of $H_0$: one galaxy each for \texttt{S250207bg} and \texttt{GW190814}, and four galaxies for \texttt{S250830bp}. For all events, we find the probability of at least one galaxy being a random association to be 29--36\%.

Publicly available databases do not provide clear evidence that these galaxies are members of identified galaxy clusters. 
Furthermore, the identification of the only potential host for \texttt{S250207bg} as an AGN may be suggestive, since the fraction of massive galaxies hosting an AGN is expected to be $\lesssim25\%$ in the local universe (e.g., \cite{Sabater_2019, pucha2025triplingcensusdwarfagn}). We therefore encourage further follow-up studies and observations of the identified galaxies and their surrounding environments to better characterize these systems and search for possible fainter structures.

The photometric bands in which the candidate galaxies are identified may also be astrophysically informative. Any systematic preference for candidate hosts to appear in a given band could indicate which galactic property is more closely linked to the sites of CBCs. In particular, redder bands are expected to more closely trace stellar mass, whereas bluer bands are more sensitive to star formation rate. By performing the search across multiple photometric bands, our method allows us to probe these different regimes.

We find that the identified galaxies for \texttt{S250207bg} and \texttt{S250830bp} are all found in the bluer $B$ and $B_J$ bands, while the galaxy associated with \texttt{GW190814} is found in the redder $J$ band. {The substantial difference in the component masses of \texttt{S250207bg} ($(m_1,m_2)=(35.2^{+1.7}_{-1.7},30.6^{+1.5}_{-1.8})M_\odot$; \cite{S250207bg_2026}), and \texttt{GW190814} ($(m_1,m_2)=(23.2^{+1.1}_{-1.0},2.59^{+0.08}_{-0.09})M_\odot$; \cite{Abbott_GW190814}), together with their association with distinct photometric bands, may suggest that CBCs with masses trace different galaxy properties.}

It is also possible that CBCs preferentially occur in small, low-luminosity galaxies. If so, we would expect to systematically find no plausible host candidates across events. The identification of 1, 1, and 4 candidates for the three events, respectively, may already provide tentative evidence against this scenario.

Because our method focuses on the most luminous galaxies, it is insensitive to catalog incompleteness, which primarily affects the faint end of the galaxy distribution. We therefore do not expect our results to change with a more complete galaxy catalog.

The number of most luminous galaxies considered for each event, $N_g$, affects both the probability of true association and the chance of random association. As $N_g$ increases, the probability of including the true host increases, but so does the chance of a random association. Assuming that the probability of hosting CBCs scales with galaxy luminosity, the gain in true association probability diminishes as progressively fainter galaxies are included, making it inefficient to increase $N_g$ further. In this paper, we adopt $\rho_g = 10^{-2}\ \mathrm{Mpc}^{-3}$ and $f_g = 1\%$ to determine $N_g$. 
As more well-localized events reveal systematic preferences for particular galaxy properties, we will gain a clearer understanding of how galaxy luminosity relates to the probability of hosting CBCs and refine these choices accordingly.

We expect the chance of random association to decrease significantly as the GW detector network sensitivity improves. For example, in the fifth LVK observing run, the detector sensitivity is expected to improve by roughly a factor of 1.7 \citep[]{Prospects}.
The localization volume of events similar to \texttt{S250207bg} is expected to be reduced by a factor of 1.7$^3$, and the chance of random association will be reduced to $< 16\%$.

\texttt{GW170817} \citep[]{GW170817_Observation} is a compelling example for future GW events with exceptionally small localization volumes. Its $V_{\rm C} \approx 290\,\mathrm{Mpc}^3$ implies $N_g < 1$ under our method. We therefore increase $N_g$ to boost the probability of true association and find that, for $N_g = 4$, one galaxy falls within the consistent $H_0$ range. This galaxy, \textsc{NGC~4968}, belongs to the same cluster as \textsc{NGC~4993}, the true host of \texttt{GW170817}. This test demonstrates that our method indeed recovers galaxies belonging to the cluster of the true host.

Assuming the identified galaxies are the true hosts or belong to the same clusters as the true hosts, we report the $H_0$ posteriors in Table~\ref{tab:H0_values}. We note that incorrect host associations are more likely to bias $H_0$ toward higher values. This is because there are more galaxies in the larger volume at higher redshift for any given LOS, leading to a higher chance of random association. 
Such systematic effect on $H_0$ could be mitigated by incorporating the probability of random association into the analysis, which would broaden and lower the $H_0$ posteriors reported in Table~\ref{tab:H0_values}. We defer this to our future work focusing on the standard siren aspect.

We also note that more accurate corrections for GW selection effects, when combining $H_0$ measurements from multiple events, will become available once LVK releases detailed information for \texttt{S250207bg} and \texttt{S250830bp}. At present, we adopt a simplified correction assuming Euclidean geometry, without incorporating the nature of the events (e.g., their masses) or the detailed detector sensitivity. In reality, a more accurate treatment assigns greater weight to higher $H_0$ values relative to this simplified approach, reflecting deviations from Euclidean geometry due to cosmic expansion (see the Appendix). Consequently, the more accurate corrections for GW selection effects are expected to systematically increase the $H_0$ values reported in Table~\ref{tab:H0_values}.

Although the $H_0$ values reported in Table~\ref{tab:H0_values} show a $1\text{--}2\sigma$ tension with the Planck measurement~\citep{Planck:2018vyg}, we refrain from drawing definitive conclusions at this stage, as the aforementioned systematic uncertainties must first be properly addressed. We will wait for the GW event information released by the LVK Collaboration to conduct a robust $H_0$ measurement. 

\section*{Acknowledgements}
The authors would like to thank Aasim Jan for the LIGO Scientific Collaboration internal review, and Aaron Zimmerman, Yixuan Dang, Archisman Ghosh, Surhud More, Konstantin Leyde, Utkarsh Mali, Grégoire Pierra, Zhuotao Li, and Fabrizio Gentile for useful discussions and comments.
A.S. and H.-Y. C. are supported by the National Science Foundation Grant PHY-2308752 and Department of Energy Grant DE-SC0025296. DEH was supported by NSF grant PHY-2513312, NSF-Simons AI-Institute for the Sky (SkAI) via grants NSF AST-2421845 and Simons Foundation MPS-AI-00010513, and by the Simons Collaboration on Black Holes and Strong Gravity through grant SFI-MPS-BH-00012593-07. The authors are grateful for computational resources provided by the LIGO Laboratory and supported by National Science Foundation Grants PHY-0757058 and PHY-0823459. This is LIGO Document P2600182-v2. This work made use of \texttt{numpy} \cite{2020NumPy-Array}, \texttt{matplotlib} \cite{matplotlib}, \texttt{astropy} \cite{The_Astropy_Collaboration_2022}, \texttt{healpy} \cite{Zonca2019}, and \texttt{ligo.skymap} \cite{Singer_2016a, Singer_2016}.

\appendix

\section{Sky map analysis}
\label{app:skymap_analysis}
GraceDB sky maps describe the GW localization as a pixelated probability distribution on the sky, and are built through the \texttt{ligo.skymap} package \citep[]{Singer_2016a, Singer_2016}. Each pixel corresponds to a finite solid-angle element, approximated as a single LOS centered at $(\alpha,\delta)$, and is assigned a probability density $p(\alpha, \delta)$. Along each LOS, the luminosity-distance posterior is summarized by a triplet $(\mu_{D_{\rm L}}, \sigma_{D_{\rm L}}, n_{D_{\rm L}})$, representing the mean, standard deviation, and normalization constant, respectively. For the $i$-th LOS $(\alpha_i, \delta_i)$, the conditional luminosity-distance posterior is modeled as \citep[]{Singer_2016}
\begin{equation}
p(D_{\rm L} | \mathcal{D}, \alpha_i, \delta_i)
= D_{\rm L}^2n^i_{D_{\rm L}}
\mathcal{N}\left(\mu^i_{D_{\rm L}}, \sigma^i_{D_{\rm L}}\right),
\end{equation}
where $\mathcal{N}(\mu, \sigma)$ is a Normal distribution centered at $\mu$ and with standard deviation $\sigma$.

The refinement level of a sky map, and thus the number of LOSs, can vary among different GW events, but it can be modified using the tools provided by the \texttt{healpy} package. For uniformity and computational convenience, we reproject all sky maps to a common resolution corresponding to
$N_{\rm LOS} = 12\ \texttt{nside}^2$, with \texttt{nside}$=1024$, yielding equal-area pixels of area $A = 9.99\times 10^{-7}\,{\rm sr}$
(or equivalently $A_{\rm deg} = 3.28\times 10^{-3}\,{\rm deg}^2$).

We first select only the LOSs lying within the 90\% credible region of the sky localization. For each selected LOS, we then compute the symmetric 90\% credible interval of the conditional luminosity-distance posterior, obtaining a lower and upper bound $(D_{\rm L}^-, D_{\rm L}^+)$.
Each luminosity-distance interval is then converted into a comoving-distance interval $[D_{\rm C}^-, D_{\rm C}^+]$, adopting our fiducial cosmology.
Finally, we compute the total comoving localization volume of each event as
\begin{equation}
    V_{\rm C}
    = A \sum_{i}^{N_{\rm LOS}}
    \frac{1}{3}
    \left[
    \left(D_{{\rm C},i}^+\right)^3
    -
    \left(D_{{\rm C},i}^-\right)^3
    \right],
\end{equation}
where the sum is restricted to LOSs within the 90\% credible region.

\section{Galaxies selection}
\label{app:galaxies_selection}
Since \textsc{GLADE+} reports the redshift of each galaxy, in addition to its sky coordinates, we first convert the luminosity-distance extension to a redshift extension of the GW event.
Given the intrinsic uncertainties in both galaxy redshift measurements and GW luminosity-distance estimates, as well as the current lack of precise measurements of the true value of the Hubble constant, we adopt a conservative approach when defining the redshift extent of the localization volume.
Specifically, we consider a broad range of Hubble constant values $[H_0^{\rm low}, H_0^{\rm high}] = [50, 140]\ \mathrm{km\,s^{-1}\,Mpc^{-1}}$,
which encompasses all recent observational constraints. 
For each LOS $i$, we convert the luminosity-distance bounds $(D_{{\rm L},i}^-, D_{{\rm L},i}^+)$ into an extended redshift interval
\begin{equation}
    [z_i^-, z_i^+] =
    \left[
    z\!\left(D_{{\rm L},i}^-, H_0^{\rm low}\right),
    z\!\left(D_{{\rm L},i}^+, H_0^{\rm high}\right)
    \right],
\end{equation}
where $z(D_{\rm L}, H_0)$ denotes the redshift corresponding to a luminosity-distance $D_{\rm L}$ for a given value of the Hubble constant, assuming a flat $\Lambda$CDM cosmology.
Finally, from the \textsc{GLADE+} catalog we select all galaxies whose sky position $(\alpha, \delta)$ falls within the 90\% sky credible region of the event, and whose redshift satisfies $z \in [z_i^-, z_i^+]$ for the corresponding LOS, thereby defining the set of galaxies consistent with the GW localization volume.

From the original set of the five most well-localized GW events, we discarded \texttt{S240925n} and \texttt{S241011k}, as the spatial distribution of galaxies within their localization volumes is strongly affected by their position on the sky.

Specifically, \texttt{S240925n} lies almost entirely along the Galactic plane, where extinction results in a highly sparse galaxy catalog, effectively preventing a meaningful host-environment analysis. 
The event \texttt{S241011k} is only partially overlapping with the Galactic plane, leading to an uneven and incomplete coverage of its sky localization region.

\begin{figure}[h!]
    \centering
    \includegraphics[width=1\linewidth]{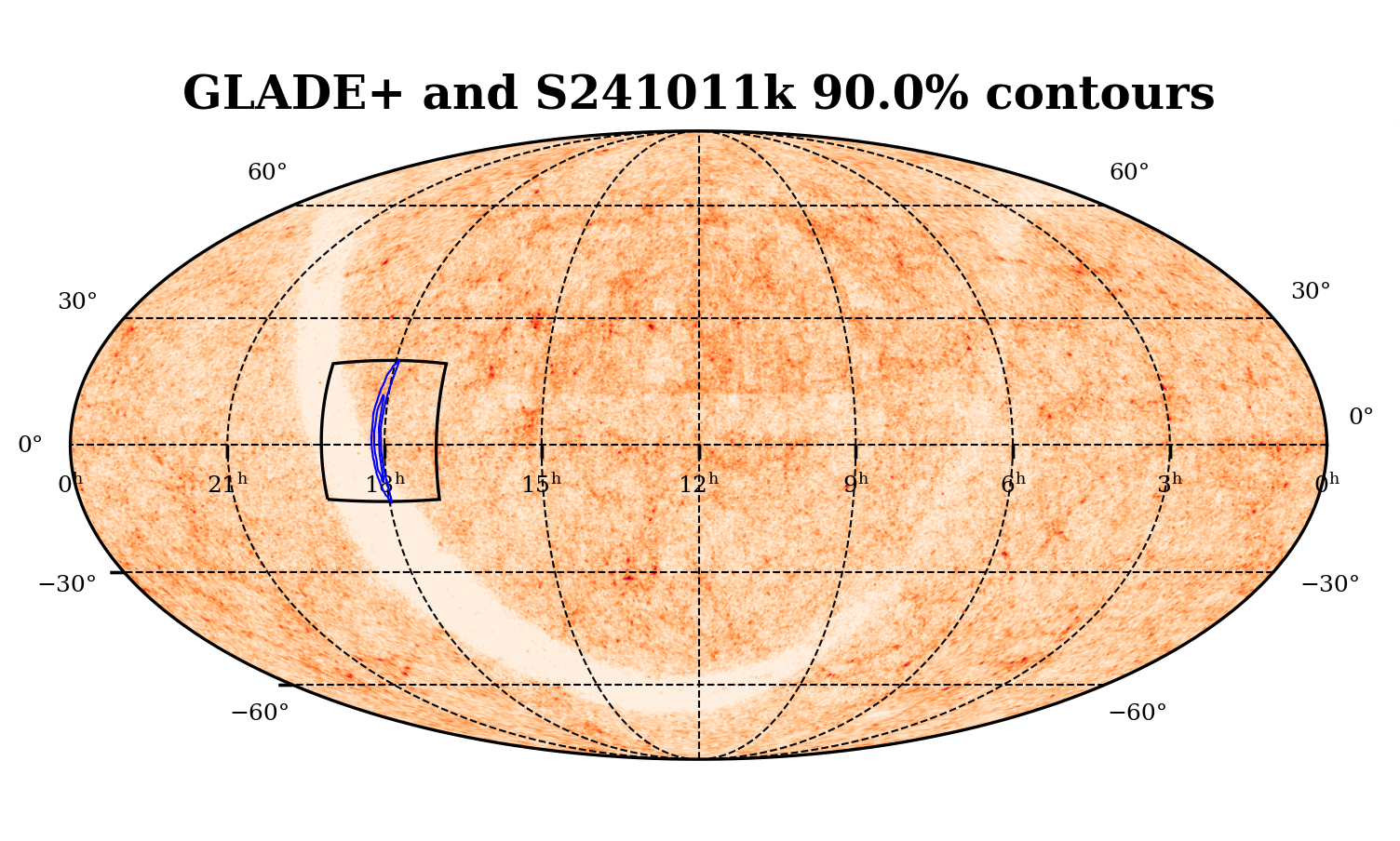}
    \includegraphics[width=0.8\linewidth]{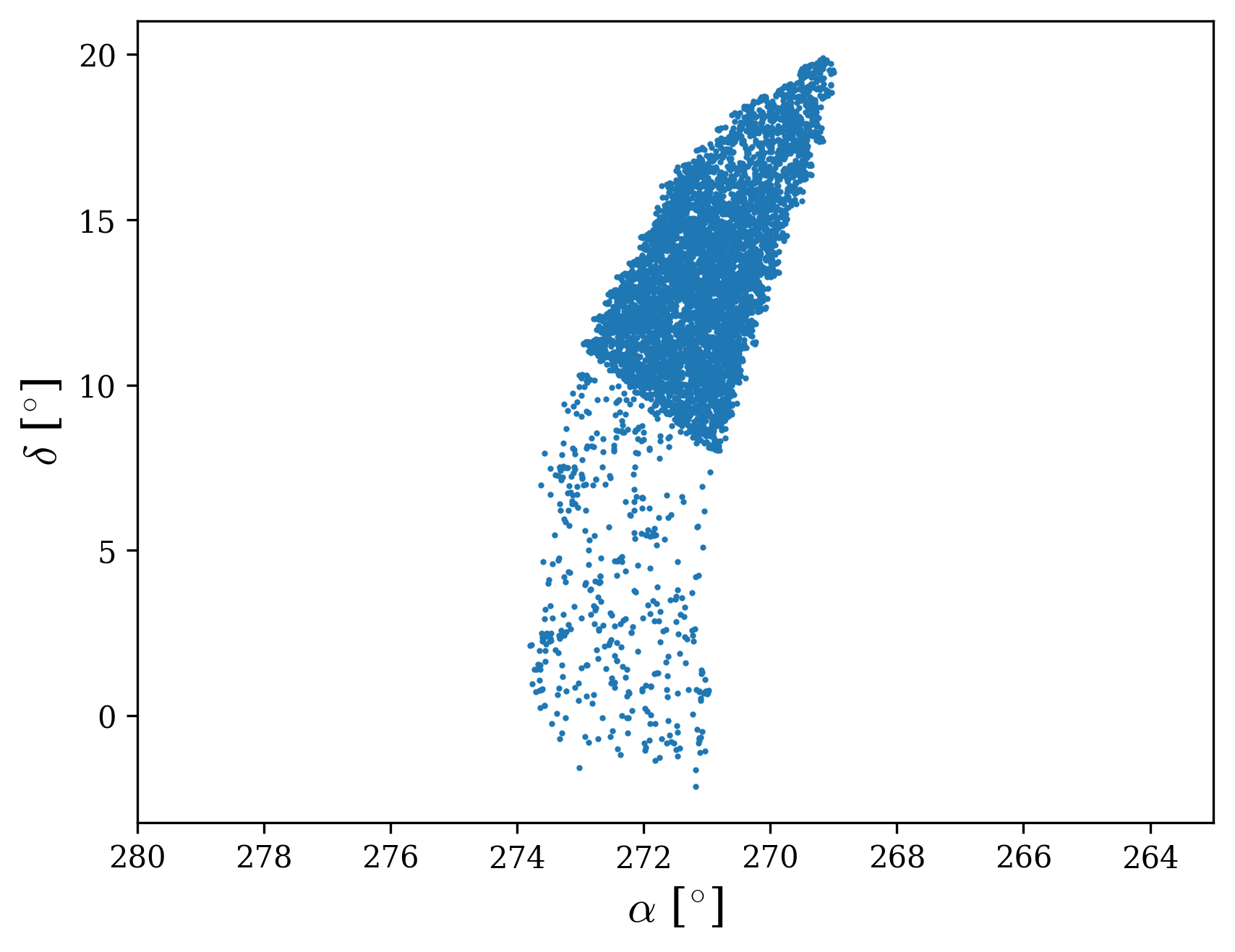}
    \caption{Top panel: 90\% sky map credible region for \texttt{S241011k} and \textsc{GLADE+} galaxies (red dots). Bottom panel: spatial distribution of \textsc{GLADE+} galaxies within the localization volume.}
    \label{fig_app:S241011k_galaxies_spatial}
\end{figure}

An example of this effect is shown in Figure \ref{fig_app:S241011k_galaxies_spatial}.

\section{$H_0$ posterior}
\label{app:posterior}
We compute the $H_0$ posterior starting from the Bayes' theorem \citep[]{Bayes1763, jeffreys1998theory}:
\begin{equation}
    p(H_0 | \mathcal{D})
    = \pi(H_0)\,
    \frac{\mathcal{L}(\mathcal{D} | H_0)}{p(\mathcal{D})},
    \label{eq:bayes_h0}
\end{equation}
where $\pi(H_0)$ denotes the prior on the Hubble constant, $\mathcal{L}(\mathcal{D} | H_0)$ the likelihood, and $p(\mathcal{D})$ the Bayesian evidence, while $p(H_0 | \mathcal{D})$ is the posterior probability distribution.

In the statistical dark siren method, a GW event is not assigned to a single host galaxy. Instead, all galaxies along each LOS within the GW localization volume are treated as possible hosts. From Eq. \ref{eq:bayes_h0}, we can rewrite the likelihood $\mathcal{L}(\mathcal{D}|H_0)$ as 
\begin{equation}
    \mathcal{L}(\mathcal{D}|H_0) = \frac{1}{\beta{(H_0)}}\int{\rm d}D_{\rm L}{\rm d}z{\rm d}\alpha{\rm d}\delta\, \mathcal{L}(\mathcal{D}, D_{\rm L}, z, \alpha, \delta|H_0),
    \label{eq:initial_likelihood}
\end{equation}
where we include the selection function
\begin{equation}
    \beta(H_0) = \int_{\mathcal{D}>\mathcal{D_{\rm th}}} {\rm d}\mathcal{D}\int{\rm d}D_{\rm L}{\rm d}z{\rm d}\alpha{\rm d}\delta\, \mathcal{L}(\mathcal{D}, D_{\rm L}, z, \alpha, \delta|H_0),
\end{equation}
with $\mathcal{D}_{\rm th}$ the detection threshold of GW detectors.

By applying the product rule of joint probabilities (e.g., \citep{Chen_2018, Salvarese:2025qel}), Eq.~\ref{eq:initial_likelihood} can be rewritten as
\begin{equation}
\begin{split}
    \mathcal{L}(\mathcal{D}|H_0)
    \propto \frac{1}{\beta(H_0)}
    \int&\mathrm{d}\alpha\mathrm{d}\delta\mathrm{d}z\;
    \mathcal{L}\!\left(\mathcal{D} | D_{\rm L}(z, H_0), \alpha, \delta \right)\\ 
    &\times p(\alpha, \delta)\,p_{\rm CBC}(z | \{z\}_{\alpha,\delta}),
\end{split}
\end{equation}
where $\mathcal{L}(\mathcal{D}|D_{\rm L}(z, H_0), \alpha, \delta)$ is the three-dimensional likelihood of the GW event, $D_{\rm L}(z, H_0)$ is the luminosity-distance corresponding to redshift $z$ for a given value of the Hubble constant in a flat $\Lambda$CDM cosmology, and $\{z\}_{\alpha,\delta}$ denotes the set of redshifts of the galaxies along the $(\alpha, \delta)$ LOS.

Since we are effectively considering a pixelized sky map, the integral over the sky coordinates can be written as a discrete sum:
\begin{equation}
    \begin{split}
        \mathcal{L}(\mathcal{D}|H_0)
        \propto \frac{1}{\beta{(H_0)}}
        \sum_{i=1}^{N_{\rm LOS}}\bigg[\int&\mathrm{d}z 
        \mathcal{L}\left(\mathcal{D} | D_{\rm L}(z, H_0), \alpha_i, \delta_i \right)\\ 
        &\times p_{\rm CBC}(z | \{z\}_i)\bigg]\,p(\alpha_i, \delta_i).
    \end{split}
    \label{eq:ds_likelihood}
\end{equation}

We now model $p_{\rm CBC}(z|\{z\}_i)$ as \citep{Gair_2023}
\begin{equation}
    p_{\rm CBC}(z|\{z\}_i) = \frac{\mathcal{L}(z|\{z\}_i)p(z)}{\int {\rm d}z\,\mathcal{L}(z|\{z\}_i)p(z)}.
\end{equation}
Here, $p(z)$ is the redshift prior on the background galaxy distribution, which we set to be uniform in comoving volume:
\begin{equation}
    p(z) \propto \frac{{\rm d}V_{\rm C}}{{\rm d}z}.
\end{equation}
For CBC events, this prior is weighted by an additional factor $1/(1+z)$ to convert the source-frame merger rate into the corresponding detector-frame rate.
For $\mathcal{L}(z|\{z\}_i)$, we assume
\begin{equation}
    \mathcal{L}(z|\{z\}_i) = \frac{1}{N_{\rm gal}^i}\sum_{j=1}^{N^i_{\rm gal}}\mathcal{N}(z|z_j, \sigma_{z_j}),
\end{equation}\\
with $N_{\rm gal}^i$ the number of galaxies along the $i$-th LOS and $\mathcal{N}(z|z_{j}, \sigma_{z_j})$ a normal distribution with mean and standard deviation the observed redshift $z_j$ and its measured error $\sigma_{z_j}$, respectively.

In the case where we are considering only one galaxy, identified by the triplet $(\alpha_*, \delta_*, z_*)$, Eq. \ref{eq:ds_likelihood} reduces to
\begin{equation} 
    \begin{split} 
        \mathcal{L}(\mathcal{D}|H_0) \propto\frac{1}{\beta{(H_0)}}\int &{\rm d}z\,\mathcal{L}(\mathcal{D}|D_{\rm L}(z, H_0),\alpha_*,\delta_*)\\ 
        &\times\mathcal{N}(z|z_*, \sigma_{z_*})\frac{1}{1+z}\frac{{\rm d}V_{\rm C}}{{\rm d}z}.
        \label{eq:bs_likelihood} 
    \end{split} 
\end{equation}
The luminosity-distance likelihood along the LOS $(\alpha_*, \delta_*)$ is obtained by dividing the conditional $D_{\rm L}$ posterior provided in the sky map by the  $D_{\rm L}$ prior adopted in the parameter-estimation (PE) pipeline. For events analyzed with \texttt{Bilby} (\texttt{S250207bg} and \texttt{S250830bp}; \cite{bilby_paper}), we assume a prior uniform in comoving volume in $D_{\rm L}$, computed using the Planck 2015 cosmology assumed in the PE pipeline (consistent with the fiducial cosmology introduced above). For events analyzed with \texttt{LALInference} (\texttt{GW190814}; \cite{lalsuite}) we use a luminosity-distance prior $\pi(D_{\rm L}) \propto D_{\rm L}^2$.

Since all the events considered in this work are located in the local universe, we adopt the approximation $\beta(H_0)\propto H_0^3$ \citep[]{Chen_2018, Gair_2023}. This scaling follows from the low-redshift Euclidean limit, in which the differential comoving volume reduces to ${\rm d}V_{\rm C}/{\rm d}z\propto z^2$, with $z\approx D_{\rm L}H_0/c$. At higher redshift, cosmological expansion introduces deviations from this approximate scaling.

\bibliography{apssamp}

\end{document}